\documentclass{article}
\usepackage{fullpage}

\newif\ifEAfigs
\EAfigstrue

\newif\ifRuleLines
\RuleLinestrue

\newlength{\RuleWidth}
\newcommand{\BeginRule}{
\ifEAfigs \begin{figure}[htbp] 
\fi
\begin{center}
\ifRuleLines \rule{\RuleWidth}{.01in} \\ \fi
\begin{minipage}[t]{\RuleWidth}
\begin{em}
\begin{tabbing}
mmm\=mmm\=mmm\=mmm\=mmm\=mmm\=mmm\=mmm\=mmm\=mmm\=\kill
}

\newcommand{\EndRule}[2]{
\end{tabbing}
\end{em}
\end{minipage}
\ifRuleLines \rule{\RuleWidth}{.01in} \fi
\end{center}
\ifEAfigs  \caption{\label{#1} #2}
          \end{figure}  
\fi
} 

\newcommand{\FigRules}{\EAfigstrue}

\newcommand{\RuleLines}{\RuleLinestrue}

\newcommand{\Abbreviation}[1]{{\sf Abbreviation: #1}\\ }

\newcommand{\If}{{\bf if\ }}
\newcommand{\Then}{{\bf then\ }}
\newcommand{\Else}{{\bf else\ }}
\newcommand{\Elseif}{{\bf elseif\ }}
\newcommand{\Endif}{{\bf endif\ }}

\newcommand{\And}{{\bf and\ }}
\newcommand{\Or}{{\bf or\ }}

\newcommand{\Choose}{{\bf choose\ }}

\newcommand{\Satisfying}{{\bf satisfying\ }}

\newcommand{\Endchoose}{{\bf endchoose\ }}

\newcommand{\vbl}[1]{{\sf #1\/}}
\newcounter{problem}
\newenvironment{problem}{\begin{trivlist}\refstepcounter{problem}%
  \item[]{\bf Problem \arabic{problem}}}{\end{trivlist}}

\title{Broy-Lamport Specification Problem:\\
A Gurevich Abstract State Machine Solution%
\thanks{University of Michigan EECS Department
Technical Report CSE-TR-320-96.}}
\author{James K. Huggins%
\thanks{\texttt{huggins@eecs.umich.edu}.  Partially supported by ONR grant
N00014-94-1-1182 and NSF grant CCR-95-04375.}
\\
EECS Department, University of Michigan, Ann Arbor, MI, 48109-2122, USA.}

\newtheorem{lem}{Lemma}
\newtheorem{thm}{Theorem}

\begin{document}
\maketitle

\begin{abstract}
We apply the Gurevich Abstract State Machine methodology to 
a benchmark specification problem of Broy and Lamport.
\end{abstract}

%%%%%%%%%%%%%%%%%%%%%%%%%%%%%%%%%%%%%%%%%%%%%%%%%%%%%%%%%%%%%%%%%%%%%%

As part of the Dagstuhl Workshop on Reactive Systems, Manfred Broy and
Leslie Lamport proposed a ``Specification Problem'' \cite{specprob}.
The problem calls for the specification and validation of a small
distributed system dealing with a remote procedure call interface.
Broy and Lamport invited proponents of different formal methods to
specify and validate the system, in order to compare the results of
different methods on a common problem.

We take up the challenge and specify the problem using the Gurevich
abstract state machine (ASM) methodology.  This paper is
self-contained.  In Section 1, we present an introduction to Gurevich
abstract state machines, including real-time machines.  The remaining
sections contain the original problem description of Broy and Lamport,
interspersed with our ASM specifications and validations.

%%%%%%%%%%%%%%%%%%%%%%%%%%%%%%%%%%%%%%%%%%%%%%%%%%%%%%%%%%%%%%%%%%%%%%

\paragraph*{Acknowledgements.}  The suggestion to give a ASM
solution to this problem was made by Egon B\"orger and
Yuri Gurevich; in particular, Yuri Gurevich actively contributed
to an early version of the work \cite{bltech}.  Leslie Lamport
made comments on an early draft of this work; Chuck Wallace made
comments on a later draft.

%%%%%%%%%%%%%%%%%%%%%%%%%%%%%%%%%%%%%%%%%%%%%%%%%%%%%%%%%%%%%%%%%%%%%%

\section{Gurevich Abstract State Machines}

Gurevich abstract state machines, formerly known as
\emph{evolving algebras} or \emph{ealgebras}, were introduced
in \cite{tutorial}; a more
complete definition (including distributed aspects) appeared
in \cite{guide}.  A discussion of real-time ASMs
appeared most recently in \cite{railroad}.

We present here a self-contained introduction to ASMs.
Sections \ref{beginea} through \ref{endea} describe distributed ASMs
(adapted from \cite{equiv}); section \ref{realea} describes
real-time ASMs.  Those already familiar with ASMs
may skip ahead to section~\ref{startbroylamp}. 

\subsection{States}
\label{beginea}
The states of a ASM are simply the structures of
first-order logic, except that relations are treated as Boolean-valued
functions.  

A \emph{vocabulary} is a finite collection of function names, each
with a fixed arity.  Every ASM vocabulary contains the
following \emph{logic symbols}: nullary function names \emph{true},
\emph{false}, \emph{undef}, the equality sign, (the names of) the usual
Boolean operations, and (for convenience) a unary function name Bool.
Some function symbols (such as Bool) are tagged as \emph{relations};
others may be tagged as \emph{external}.

A \emph{state} $S$ of vocabulary $\Upsilon$ is a non-empty set $X$ (the
\emph{superuniverse} of $S$), together with interpretations of all
function symbols in $\Upsilon$ over $X$ (the \emph{basic functions} of
$S$).  A function symbol $f$ of arity $r$ is interpreted as an $r$-ary
operation over $X$; if $r=0$, $f$ is interpreted as an element of $X$.
The interpretations of the function symbols \emph{true}, \emph{false},
and \emph{undef} are distinct, and are operated upon by the Boolean
operations in the usual way. 

Let $f$ be a relation symbol of arity $r$.  We require that (the
interpretation of) $f$ is \emph{true} or \emph{false} for every
$r$-tuple of elements of $S$.  If $f$ is unary, it can be viewed as a
\emph{universe}: the set of elements $a$ for which $f(a)$ evaluates to
\emph{true}.  For example, \emph{Bool} is a universe consisting of
the two elements (named) \emph{true} and \emph{false}.

Let $f$ be an $r$-ary basic function and $U_0, \ldots, U_r$ be
universes.  We say that $f$ has \emph{type} $U_1 \times \ldots \times
U_r \rightarrow U_0$ in a given state if $f(\bar{x})$ is in
the universe $U_0$ for every $\bar{x} \in U_1 \times \ldots \times
U_r$, and $f(\bar{x})$ has the value \emph{undef} otherwise.

\subsection{Updates}

The simplest change that can occur to a state is the change of
an interpretation of one function at one particular tuple of arguments.
We formalize this notion.

A \emph{location} of a state $S$ is a pair $\ell = (f,\bar{x})$, where
$f$ is an $r$-ary function name in the vocabulary of $S$ and $\bar{x}$
is an $r$-tuple of elements of (the superuniverse of) $S$.  (If $f$ is
nullary, $\ell$ is simply $f$.)  An \emph{update} of a state $S$
is a pair $(\ell,y)$, where $\ell$ is a location of $S$ and $y$ is
an element of $S$.  To \emph{fire} $\alpha$ at $S$, put $y$ into
location $\ell$; that is, if $\ell = (f,\bar{x})$, redefine $S$
to interpret $f(\bar{x})$ as $y$ and leave everything else unchanged.

\subsection{Transition Rules}
\EAfigsfalse
\RuleLinesfalse

We introduce rules for describing changes to states.  At a given state
$S$ whose vocabulary includes that of a rule $R$, $R$ gives rise to a
set of updates; to execute $R$ at $S$, fire all the updates in the
corresponding update set.  We suppose throughout that a state of
discourse $S$ as a sufficiently rich vocabulary.

An \emph{update instruction} $R$ has the form 
\BeginRule 
$f(t_1, t_2, \ldots, t_n$) := $t_0$ 
\EndRule{}{} 
where $f$ is an $r$-ary function name and each $t_i$ is a term.  (If
$r=0$, we write $f := t_0$ rather than $f() := t_0$.) The update set for
$R$ contains a single update $(\ell,y)$, where $y$ is the value
$(t_0)_S$ of $t_0$ at $S$, and $\ell = (f, (x_1, \ldots, x_r))$, where
$x_i = (t_i)_S$.  In other words, to execute $R$ at $S$, set $f(x_1,
\ldots, x_n)$ to $y$, where $x_i$ is the value of $t_i$ at $S$ and $y$
is the value of $t_0$ at $S$. 

A \emph{block rule} $R$ is a sequence $R_1, \ldots, R_n$ of transition
rules.  To execute $R$ at $S$, execute all the $R_i$ at $S$
simultaneously.  That is, the update set of $R$ at $S$ is the union of
the update sets of the $R_i$ at $S$.

A \emph{conditional rule} $R$ has the form
\BeginRule
\If $g$ \Then $R_0$ \Else $R_1$ \Endif
\EndRule{}{}
where $g$ (the \emph{guard}) is a term and $R_0, R_1$ are rules.
The meaning of $R$ is the obvious one: if $g$ evaluates to \emph{true}
in $S$, then the update set for $R$ at $S$ is the same as that for
$R_0$ at $S$; otherwise, the update set for $R$ at $S$ is the same as
that for $R_1$ at $S$.

A \emph{choice rule} $R$ has the form
\BeginRule
\Choose $v$ \Satisfying c(v)\\
   \>$R_0(v)$\\
\Endchoose
\EndRule{}{}
where $v$ is a variable, $c(v)$ is a term involving variable $v$, and
$R_0(v)$ is a rule with free variable $v$.  This rule is
nondeterministic.  To execute $R$ in state $S$, choose some element of
$a$ of $S$ such that $c(a)$ evaluates to \emph{true} in $S$, and execute
rule $R_0$, interpreting $v$ as $a$.  If no such element exists, do
nothing. 

\subsection{Distributed Machines}
\label{endea}

In this section we describe how distributed ASMs evolve over time.
The intuition is that each agent of a distributed ASM operates in a
sequential manner, and moves of different agents are ordered only when
necessary (for example, when two agents attempt contradictory
updates).

How do ASMs interact with the external world?  There are several ways
to model such interactions.  One common way is through the use of
\emph{external functions}.  The values of external functions are
provided not by the ASM itself but by some external oracle.  The value
of an external function may change from state to state without any
explicit action by the ASM.  If $S$ is a state of a ASM, let $S^-$ be
the reduct of $S$ to (the vocabulary) of non-external functions.

Let $\Upsilon$ be a vocabulary containing the universe \emph{agents},
a unary function \emph{Mod}, and a nullary function \emph{Me}.  
A \emph{distributed ASM program} $\Pi$ of vocabulary
$\Upsilon$ consists of a finite set of \emph{modules}, each of which
is a transition rule over the vocabulary $\Upsilon$.  Each module has
a unique name different from $\Upsilon$ or \emph{Me}.  The intuition
is that a module is a program to be executed by one or more agents.

A (global) \emph{state} of $\Pi$ is a structure $S$ of vocabulary
$\Upsilon-\{\emph{Me}\}$, where different module names are interpreted
as different elements of $S$ and \emph{Mod} maps module names to
elements of \emph{agents} and all other elements to \emph{undef}.  If
\emph{Mod($\alpha$) = $M$}, we say that $\alpha$ is an \emph{agent}
with program $M$.

For every agent $\alpha$, View$_\alpha(S)$ is the reduct of $S$ to the
functions mentioned in $\alpha$'s program \emph{Mod($\alpha$)},
extended by interpreting the special function \emph{Me} as $\alpha$.
View$_\alpha(S)$ can be seen as the local state of agent $\alpha$
corresponding to the global state $S$.  To \emph{fire} an agent
$\alpha$ at a state $S$, execute \emph{Mod($\alpha$)} at state
View$_\alpha(S)$.

A \emph{run} of a distributed ASM program $\Pi$ is
a triple $(M, A, \sigma)$, satisfying the following conditions:

\begin{enumerate}
\item $M$, the set of \emph{moves} of $\rho$, is a partially 
ordered set where every set $\{\nu: \nu \leq \mu\}$ is finite.  

Intuitively, $\nu < \mu$ means that move $\nu$ occurs before
move $\mu$.  If $M$ is totally ordered, we call $\rho$ a 
\emph{sequential} run.

\item $A$ assigns agents (of $S_0$) to moves such that
every non-empty set $\{\mu: A(\mu) = \alpha\}$ is linearly ordered.

Intuitively, $A(\mu)$ is the agent which performs move $\mu$;
the condition asserts that every agent acts sequentially.

\item $\sigma$ maps finite initial segments of $M$ (including
$\emptyset$) to states of $\Pi$.

Intuitively, $\sigma(X)$ is the result of performing all moves
of $X$; $\sigma(\emptyset)$ is the initial state $S_0$.  

\item (Coherence) If $\mu$ is a maximal element of a finite initial
segment $Y$ of $M$, and $X = Y - \{\mu\}$, then $\sigma(Y)^-$ is
obtained from $\sigma(X)$ by firing $A(\mu)$ at $\sigma(X)$.

\end{enumerate}

\subsection{Real-Time Machines}
\label{realea}

Real-time ASMs are an extension of distributed ASMs
which incorporate the notion of real-time.  The notion of
state is basically unchanged; what changes is the notion of run.  The
definitions presented here are taken from \cite{railroad}; they may
not be sufficient to model every real-time system but certainly
suffice for the models to be presented in this paper.

Let $\Upsilon$ be a vocabulary with a universe symbol \emph{Reals} which
does not contain the nullary function \emph{CT}.  Let $\Upsilon^+$ be
the extension of $\Upsilon$ to include \emph{CT}.  We will restrict
attention to $\Upsilon^+$-states where the universe \emph{Reals} is the
set of real numbers and \emph{CT} evaluates to a real number.
Intuitively, \emph{CT} gives the current time of a given state. 

A \emph{pre-run} $R$ of vocabulary $\Upsilon^+$ is a mapping from the
interval $[0,\infty)$ to states of vocabulary $\Upsilon^+$ satisfying
the following requirements, where $\rho(t)$ is the reduct of $R(t)$ to
$\Upsilon$:
\begin{enumerate}
\item The superuniverse of every $R(t)$ is that of $R(0)$; that is,
the superuniverse does not change during the pre-run.
\item At every $R(t)$, \emph{CT} evaluates to $t$.  \emph{CT}
represents the current (global) time of a given state.
\item For every $\tau > 0$, there is a finite sequence $0 = t_0 < t_1
< \ldots < t_n = \tau$ such that if $t_i < \alpha < \beta < t_{i+1}$,
then $\rho(\alpha) = \rho(\beta)$.  That is, for every time $t$,
there is a finite, discrete sequence of moments prior to $t$
when a change occurs in the states of the run (other than a change
to \emph{CT}).
\end{enumerate}

In the remainder of this section, let $R$ be a pre-run of 
vocabulary $\Upsilon^*$ and $\rho(t)$ be the reduct of $R(t)$ to
$\Upsilon$.  $\rho(t+)$ is any state
$\rho(t+\epsilon)$ such that $\epsilon > 0$ and $\rho(t+\delta) =
\rho(t+\epsilon)$ for all positive $\delta < \epsilon$.  Similarly, if
$t>0$, then $\rho(t-)$ is any state $\rho(t-\epsilon)$ such that $0 <
\epsilon \leq t$ and $\rho(t - \delta) = \rho(t-\epsilon)$ for all
positive $\delta < \epsilon$.

A \emph{pre-run} $R$ of vocabulary $\Upsilon^+$ is a \emph{run} of
$\Pi$ if it satisfies the following conditions:
\begin{enumerate}
\item If $\rho(t+)$ differs from $\rho(t)$ then $\rho(t+)$ is the
$\Upsilon$-reduct of the state resulting from executing some
modules $M_1,\ldots,M_k$ at $R(t)$.  All external functions have
the same values in $\rho(t)$ and $\rho(t+)$.
\item If $t>0$ and $\rho(t)$ differs from $\rho(t-)$ then they
differ only in the values of external functions.  All internal
functions have the same values in $\rho(t-)$ and $\rho(t)$.
\end{enumerate}

%%%%%%%%%%%%%%%%%%%%%%%%%%%%%%%%%%%%%%%%%%%%%%%%%%%%%%%%%%%%%%%%%%%%%%

\section{The Procedure Interface}
\label{startbroylamp}
\FigRules
\RuleLines

The Broy-Lamport specification problem begins as follows:

\begin{quotation}
The problem calls for the specification and verification of a series
of {\em components}.  Components interact with one another using a
procedure-calling interface.  One component issues a {\em call\/} to
another, and the second component responds by issuing a {\em return}.
A call is an indivisible (atomic) action that communicates a procedure
name and a list of {\em arguments\/} to the called component.  A
return is an atomic action issued in response to a call.  There are
two kinds of returns, {\em normal\/} and {\em exceptional}.  A normal
call returns a {\em value\/} (which could be a list).  An exceptional
return also returns a value, usually indicating some error condition.
An exceptional return of a value $e$ is called {\em raising exception
$e$}.  A return is issued only in response to a call.  There may be
``syntactic'' restrictions on the types of arguments and return
values.

A component may contain multiple {\em processes} that can concurrently
issue procedure calls.  More precisely, after one process issues a
call, other processes can issue calls to the same component before the
component issues a return from the first call.  A return action
communicates to the calling component the identity of the process that
issued the corresponding call.
\end{quotation}

The modules in our ASM represent components; each
component is described by one module.  The universe \emph{Components}
contains (elements representing) the modules of the system.

The agents in our ASM represent processes; just as each
component can have several processes, so a given module can belong to
several agents.  The universe \emph{agents} contains elements
representing the agents of the system.  A function \emph{Component:
agents $\rightarrow$ modules} indicates the component to which a given
process belongs\footnote{For those familiar with the Lipari Guide,
this is a renaming of the function \emph{Mod}.}.

Calls and returns are represented by the execution of transition
rules which convey the appropriate information between two processes.
The following unary functions are used to transmit this information,
where the domain of each function is the universe of \emph{agents}:
\begin{itemize}
\item \emph{CallMade}: whether or not this process made a call
while handling the current call
\item \emph{CallSender}: which process made a call most recently to this
process
\item \emph{CallName}: what procedure name was sent to this process
\item \emph{CallArgs}: what arguments were sent to this process
\item \emph{CallReply}: what type of return (normal or exceptional) was
sent to this process
\item \emph{CallReplyValue}: what return value was sent to this process
\end{itemize}

We use two macros (or abbreviations), CALL and RETURN,
which are used to make the indivisible (atomic) actions of issuing
calls and returns.  Each macro performs the task of transferring the
relevant information between the caller and callee.
The definitions of CALL and RETURN are given in
Figure \ref{CALL1}.

\BeginRule
\Abbreviation{CALL(procname, arglist, destination)}
   \>\Choose $p$ \Satisfying 
            (Component(p)=destination \And CallSender(p)=undef)\\
   \>   \>CallSender(p) := Me\\
   \>   \>CallName(p) := procname\\
   \>   \>CallArgs(p) := arglist\\
   \>   \>CallMade(Me) := true\\
   \>   \>CallReply(Me) := undef\\
   \>   \>CallReplyValue(Me) := undef\\
   \>\Endchoose\\
\\
\Abbreviation{RETURN(type, value)}
   \>CallReply(CallSender(Me)) := type\\
   \>CallReplyValue(CallSender(Me)) := value\\
   \>CallSender(Me) := undef\\
   \>CallName(Me) := undef\\
   \>CallArgs(Me) := undef\\
   \>CallMade(Me) := false
\EndRule{CALL1}{Definitions of the CALL and RETURN abbreviations.}

%%%%%%%%%%%%%%%%%%%%%%%%%%%%%%%%%%%%%%%%%%%%%%%%%%%%%%%%%%%%%%%%%%%%%%

\section{A Memory Component}

The Broy-Lamport problem calls for the specification of a memory
component.  The requirements are as follows:

\begin{quotation}
The component to be specified is a memory that maintains the contents
of a set \vbl{MemLocs} of locations.  The contents of a location is an
element of a set \vbl{MemVals}.  This component has two procedures,
described informally below.  Note that being an element of \vbl{MemLocs}
or \vbl{MemVals} is a ``semantic'' restriction, and cannot be imposed
solely by syntactic restrictions on the types of arguments.
\begin{tabbing}
{\bf Return Value } \=\kill 
{\bf Name } \> \vbl{Read}\\ 
{\bf Arguments } \> \vbl{loc} : an element of \vbl{MemLocs}\\ 
{\bf Return Value } \> an element of \vbl{MemVals}\\
 {\bf Exceptions } \> \vbl{BadArg} : argument
                        \vbl{loc} is not an element of \vbl{MemLocs}.\\
                   \> \vbl{MemFailure} : the memory cannot be read.\\
{\bf Description} \> Returns the value stored in address \vbl{loc}.
               \\[.5\baselineskip]
{\bf Name } \> \vbl{Write}\\
{\bf Arguments } \> \vbl{loc} : an element of \vbl{MemLocs}\\
                 \> \vbl{val}   : an element of \vbl{MemVals}\\
{\bf Return Value } \> some fixed value \\
{\bf Exceptions } \> \vbl{BadArg} : \= argument \vbl{loc} is not an element of 
                               \vbl{MemLocs}, or\\
            \>            \> argument \vbl{val} is not an element of 
                               \vbl{MemVals}.\\
            \> \vbl{MemFailure} : the write {\em might\/} not have succeeded.\\
{\bf Description} \> Stores the value \vbl{val} in address \vbl{loc}.
\end{tabbing}
The memory must eventually issue a return for every \vbl{Read} and
\vbl{Write} call.  

Define an {\em operation\/} to consist of a procedure call and the
corresponding return.  The operation is said to be {\em successful\/}
iff it has a normal (nonexceptional) return.  The memory behaves as if
it maintains an array of atomically read and written locations that
initially all contain the value \vbl{InitVal}, such that:
\begin{itemize}
\item An operation that raises a \vbl{BadArg} exception has no effect
on the memory.

\item Each successful $\vbl{Read}(l)$ operation performs a single
atomic read to location $l$ at some time between the call and return.

\item Each successful $\vbl{Write}(l,\,v)$ operation performs a
sequence of one or more atomic writes of value $v$ to location $l$
at some time between the call and return.

\item Each unsuccessful $\vbl{Write}(l,\,v)$ operation performs a
sequence of zero or more atomic writes of value $v$ to location $l$
at some time between the call and return.
\end{itemize}
A variant of the Memory Component is the Reliable Memory Component, in
which no \vbl{MemFailure} exceptions can be raised.
\begin{problem}
(a) Write a formal specification of the Memory component and of
the Reliable Memory component.

(b) Either prove that a Reliable Memory component is a correct
implementation of a Memory component, or explain why it should not be.

(c) If your specification of the Memory component allows an
implementation that does nothing but raise \vbl{MemFailure}
exceptions, explain why this is reasonable.
\end{problem}
\end{quotation}

\subsection{ASM Description}
As suggested by the description, our ASM has universes
\emph{MemLocs} and \emph{MemVals}, and a function \emph{Memory:
MemLocs $\rightarrow$ MemVals} which indicates the contents of
memory at a given location.  We also have the following universes and
associated functions: 
\begin{itemize}
\item \emph{procnames}: names of procedures to be called.  Includes
the distinguished elements \emph{read} and \emph{write}.
\item \emph{lists}: lists of elements in the superuniverse.
Unary functions \emph{First, Second} extract the corresponding
element from the given list.
\item \emph{returntypes}: types of returns to be issued.
Includes the distinguished elements \emph{normal} and
\emph{exception}.
\item \emph{exceptions}: types of exceptions to be issued.
Includes the distinguished elements \emph{BadArg}
and \emph{MemFailure}.
\item \emph{values}: types of return values.  Includes
the universe \emph{MemVals} as well as a distinguished element
\emph{Ok} (used for returns from successful \emph{write} operations,
where the nature of the return value is unimportant).
\end{itemize}

In addition, we use two Boolean-valued external functions,
\emph{Succeed} and \emph{Fail}.  The intuition is that \emph{Fail}
indicates when a component should unconditionally fail,
while \emph{Succeed} indicates when a component may succeed
during an attempt to write to memory.  We require that
\emph{Succeed} cannot be false forever; that is, for any
state $\sigma$ in which \emph{Succeed} is false, \emph{Succeed}
cannot be false in every successor state $\rho > \sigma$.
This ensures that every operation eventually terminates.

\subsection{Component Program}
Our ASM contains an unspecified number of agents
comprising the memory component.  The program for these agents
is shown in Figure \ref{memea}.

\BeginRule
\If CallName(Me)=read \Then\\
   \>\If MemLocs(First(CallArgs(Me)))=false \Then
                    RETURN(exception, BadArg)\\
   \>\Elseif Fail \Then RETURN (exception, MemFailure)\\
   \>\Else RETURN(normal, Memory(First(CallArgs(Me))))\\
   \>\Endif\\ 
\Elseif CallName(Me)=write \Then\\
   \>\If MemLocs(First(CallArgs(Me)))=false \Or
		MemVals(Second(CallArgs(Me)))=false \Then\\
   \>   \>RETURN(exception, BadArg)\\
   \>\Elseif Fail \Then RETURN(exception, MemFailure)\\
   \>\Else\\
   \>   \>Memory(First(CallArgs(Me))) := Second(CallArgs(Me))\\
   \>   \>\If Succeed \Then RETURN(normal, Ok) \Endif\\
   \>\Endif\\
\Endif
\EndRule{memea}{Memory component program.}

In order to make this evolving algebra complete, we need a
component which makes calls to the memory component.  Of course, we don't
wish to constrain how the memory component is called, other than
that the CALL interface is used.

Our ASM contains an unspecified number of processes
implementing a calling component, whose simple program is shown 
in Figure \ref{callerea}.

\BeginRule
\If MakeCall \Then CALL(GetName, GetArgs, MemComponent) \Endif
\EndRule{callerea}{Calling component program.}

This component uses several external functions.  \emph{MakeCall}
returns a Boolean value indicating whether a call should be
made at a given moment.  \emph{GetName} and \emph{GetArgs} supply the
procedure name and argument list to be passed to the memory component.
\emph{MemComponent} is a static (non-external) function indicating the
memory component.

We define an \emph{operation} to be the linearly-ordered sequence
of moves beginning with the execution of CALL by one 
component and ending with the execution of RETURN by the
component which received the call.  In the simplest case, this
sequence has exactly two moves (an execution of CALL followed
immediately by an execution of RETURN).

A reliable memory component is identical to a memory component
except that the external function \emph{Fail} is required to
have the value \emph{false} at all times.  Thus, the
MemFailure exception cannot be raised.

\subsection{Correctness}
We now show that the Memory and Reliable Memory components
specified above satisfy the given requirements.  The Broy-Lamport
problem does not require us to demonstrate that the specification
in fact satisfies the given requirements; nonetheless, it seems 
quite reasonable and important to do so.  

\begin{lem}
Every operation resulting in a \vbl{BadArg} exception has no effect on
the memory.
\end{lem}

\noindent \textbf{Proof.}  Observe from the component specification
above that any such operation consists of exactly two moves:
the original call to the Memory component, and the move
which raises the BadArg exception.  Observe further that the
rule executed by the Memory component to issue the BadArg 
exception neither reads nor alters the function \emph{Memory}.
QED.

\begin{lem}
Every successful $\vbl{Read}(l)$ operation performs a single 
atomic read to location $l$ at some time between the call and return.
\end{lem}

\noindent \textbf{Proof.}  Observe from the component specification
above that any such operation consists of exactly two moves:
the original call to the Memory component, and the move
which issues the successful return.  Observe further that the
rule executed by the Memory component to issue the successful
return accesses the \emph{Memory} function exactly once,
when evaluating \emph{Memory(First(CallArgs(Me)))}.  QED.

\begin{lem}
Every successful $\vbl{Write}(l,\,v)$ operation 
performs a sequence of one or more atomic writes of
value $v$ to location $l$ at some time between the call and return.
\end{lem}

\noindent \textbf{Proof.}  Observe from the component specification
above that any such operation consists of several moves:
the original call to the Memory component, and one or more
moves which write value $v$ to location $l$.  The last of these
writing moves also issues the successful return.  QED.

\begin{lem}
Every unsuccessful $\vbl{Write}(l,\,v)$ operation 
performs a sequence of zero or more atomic writes of
value $v$ to location $l$ at some time between the call and return.
\end{lem}

\noindent \textbf{Proof.}  Observe from the component specification
above that any such operation consists of several moves:
the original call to the Memory component, zero or more
moves which write value $v$ to location $l$, and
the move which returns an exception (either \emph{MemFailure}
or \emph{BadArg}).  QED.

\bigskip
\noindent We thus have immediately:
\begin{thm}
The ASM specification of the memory component 
correctly implements the requirements given for memory components.
\end{thm}

\noindent As to the other issues we are asked to consider:
\begin{itemize}
\item It is trivial to see that a reliable memory component is
a correct implementation of a memory component; all of the
proofs above apply to reliable memory components, other than the fact
that \vbl{Write} operations cannot raise \emph{MemFailure} exceptions.

\item Our specification does allow for a memory component to return
only MemFailure exceptions.  It seems reasonable to allow this
behavior; it corresponds to the real-world scenario where
a memory component is irreparable or cannot be reached through
the network.
\end{itemize}

%%%%%%%%%%%%%%%%%%%%%%%%%%%%%%%%%%%%%%%%%%%%%%%%%%%%%%%%%%%%%%%%%%%%%%

\section{The RPC Component}

The Broy-Lamport problem calls for the specification of an
RPC (for ``remote procedure call'') component.  Its description
is as follows:

\begin{quotation}
The RPC component interfaces with two environment components, a {\em
sender\/} and a {\em receiver}.  It relays procedure calls from the
sender to the receiver, and relays the return values back to the
sender.  Parameters of the component are a set \vbl{Procs} of
procedure names and a mapping \vbl{ArgNum}, where $\vbl{ArgNum}(p)$ is
the number of arguments of each procedure $p$.  The RPC component
contains a single procedure:
\begin{tabbing}
{\bf Return Value } \=\kill 
{\bf Name } \> \vbl{RemoteCall}\\ 
{\bf Arguments } \> \vbl{proc} : name of a procedure\\
                 \> \vbl{args} : list of arguments\\
{\bf Return Value } \> any value that can be returned by a call to \vbl{proc}\\
{\bf Exceptions } \> \vbl{RPCFailure} : the call failed\\
                  \> \vbl{BadCall} : \=\vbl{proc} is not a valid name 
                                        or \vbl{args} is not a \\
                  \>      \> syntactically correct list of arguments for 
                             \vbl{proc}.\\
                  \> Raises any exception raised by a call to \vbl{proc}\\
{\bf Description} \> Calls procedure \vbl{proc} with arguments
       \vbl{args}
\end{tabbing}
A call of $\vbl{RemoteCall}(\vbl{proc}, \vbl{args})$ causes the RPC
component to do one of the following:
\begin{itemize}
\item Raise a \vbl{BadCall} exception if \vbl{args} is not a 
      list of $\vbl{ArgNum}(\vbl{proc})$ arguments.
\item Issue one call to procedure \vbl{proc} with arguments \vbl{args},
      wait for the corresponding return (which the RPC
      component assumes will occur)
      and either (a)~return the value (normal or exceptional) returned by
      that call, or (b)~raise the \vbl{RPCFailure} exception.

\item Issue no procedure call, and raise the \vbl{RPCFailure} exception.
\end{itemize}
The component accepts concurrent calls of \vbl{RemoteCall} from the
sender, and can have multiple outstanding calls to the receiver.

\begin{problem}
Write a formal specification of the RPC component.
\end{problem}
\end{quotation}

\subsection{ASM Description}

We add a new distinguished element \emph{remotecall} to
the universe of \emph{procnames}, and distinguished elements
\emph{BadCall} and \emph{RPCFailure} to the universe of
\emph{exceptions}.  We use the standard universe of \emph{integers}
in conjunction with the following functions:
\begin{itemize}
\item \emph{ArgNum: procnames $\rightarrow$ integers} indicates
the number of arguments to be supplied with each procedure name
\item \emph{Length: lists $\rightarrow$ integers} returns
the length of the given argument list
\end{itemize}
A distinguished element \emph{Destination} indicates the component
to which this RPC component is supposed to forward its procedure call.

The ASM program for the RPC component is shown 
in Figure \ref{rpcea}.
\BeginRule
\If CallName(Me) = remotecall \Then\\
   \>\If Length(Second(CallArgs(Me))) $\neq$ ArgNum(First(CallArgs(Me))) \Then\\
   \>   \>RETURN(exception, BadCall)\\
   \>\Elseif CallMade(Me)=false \Then\\
   \>   \>\If Fail \Then RETURN(exception, RPCFailure)\\
   \>   \>\Else CALL(First(CallArgs(Me)),Second(CallArgs(Me)),Destination)\\
   \>   \>\Endif\\
   \>\Elseif CallReply(Me) $\neq$ undef \Then\\
   \>   \>\If Fail \Then RETURN(exception, RPCFailure)\\
   \>   \>\Else RETURN(CallReply(Me), CallReplyValue(Me))\\
   \>   \>\Endif\\
   \>\Endif\\
\Endif
\EndRule{rpcea}{RPC component program.}

To complete the specification, we need to supply a component
to call the RPC component (such as our caller component from
the previous section) and a component for the RPC component to
call (such as the memory component from the previous section).

Again, we are not asked to prove that the specification satisfies
the requirements given above; the proof is similar to that given
in the last section and is omitted.

%%%%%%%%%%%%%%%%%%%%%%%%%%%%%%%%%%%%%%%%%%%%%%%%%%%%%%%%%%%%%%%%%%%%%%

\section{Implementing The Memory Component}

The Broy-Lamport problem calls us to create a memory component using
a reliable memory component and an RPC component.  The requirements
are as follows:

\begin{quotation}
A Memory component is implemented by combining an RPC component with a
Reliable Memory component as follows.  A \vbl{Read} or \vbl{Write}
call is forwarded to the Reliable Memory by issuing the appropriate
call to the RPC component.  If this call returns without raising an
\vbl{RPCFailure} exception, the value returned is returned to the
caller.  (An exceptional return causes an exception to be raised.)  If
the call raises an \vbl{RPCFailure} exception, then the implementation
may either reissue the call to the RPC component or raise a
\vbl{MemFailure} exception.  The RPC call can be retried arbitrarily
many times because of \vbl{RPCFailure} exceptions, but a return from
the \vbl{Read} or \vbl{Write} call must eventually be issued.

\begin{problem}
Write a formal specification of the implementation, and prove
that it correctly implements the specification of the Memory
component of Problem 1.
\end{problem}
\end{quotation}

Our implementation includes three modules.  Two of the modules are,
naturally, instances of the reliable memory component and the
RPC component.  The program for the third module is shown
in Figure \ref{imp1ea}.

\BeginRule
\If CallName(Me) $\neq$ undef \Then\\
   \>\If CallMade(Me) = false \Then\\
   \>   \>CALL(CallName(Me), CallArgs(Me), RPCComponent)\\ 
   \>\Elseif CallReply(Me) $\neq$ undef \Then\\
   \>   \>\If (CallReply(Me) $\neq$ exception) \Or 
			(CallReplyValue(Me) $\neq$ RPCFailure) \Then\\
   \>   \>   \>RETURN(CallReply(Me),CallReplyValue(Me))\\
   \>   \>\Elseif Retry \Then 
			CALL(CallName(Me), CallArgs(Me), RPCComponent)\\ 
   \>   \>\Else RETURN(exception, MemFail)\\
   \>   \>\Endif\\
   \>\Endif\\
\Endif
\EndRule{imp1ea}{Implementing component program.}

The component uses an external Boolean-valued function \emph{Retry},
which indicates whether or not an \emph{RPCFailure} exception should result
in another attempt to send the call to the RPC component.  The
distinguished element \emph{RPCComponent} indicates the RPC component
module; the distinguished element \emph{MemFail} is a member of the
universe of \emph{exceptions}.  We require that \emph{Retry}
cannot force the component to resend the call forever; more precisely,
for any given agent, and for any state $\sigma$ such that
\emph{CallReply(Me)=exception}, \emph{CallReplyValue(Me)=RPCFailure},
and \emph{Retry=true}, not every successor state $\rho > \sigma$ which
satisfies \emph{CallReply(Me)=exception} and
\emph{CallReplyValue(Me)=RPCFailure} also satisfies \emph{Reply=true}.

It remains to prove that this implementation is correct.  We consider
the four original requirements for memory components.

\begin{lem}
Every operation resulting in a \vbl{BadArg} exception has no effect on
the memory.
\end{lem}

\noindent \textbf{Proof.}  From the module specifications given above,
we observe that an operation resulting in a \vbl{BadArg} exception
consists of the following sequence of moves:
\begin{itemize}
\item a call from the caller component to the implementing component
given above
\item a call from the implementing component to the RPC component
\item a return of the \vbl{BadArg} exception from the RPC component to
the implementing component
\item a return of the \vbl{BadArg} exception from the implementing
component to the caller component
\end{itemize}
An examination of the rules involved shows that the \emph{Memory}
function is neither read nor updated in any of these moves.  QED.

\begin{lem}
Every unsuccessful $\vbl{Read}(l)$ operation performs zero or more
atomic reads to location $l$ at some time between the call and return.
\end{lem}
This is not one of the original requirements, but the result is
used later.

\smallskip
\noindent \textbf{Proof.}  Fix a sequence of moves which comprise an
unsuccessful $\vbl{Read}$ operation.  The first element of this
sequence is the call from the caller component to the
implementation component; the last element of this sequence is the
corresponding exceptional return. 

The moves between these two elements can be divided into
one or more disjoint subsequences of moves, each of which 
is an operation of the RPC component resulting in an exceptional
return.  There are several cases.
\begin{itemize}
\item The RPC component may raise an \emph{RPCException} without
calling the reliable memory component.  In this case, 
\emph{Memory} is never accessed.
\item The RPC component may make a call to the reliable memory
component, ignore the return value, and raise an \emph{RPCException}.
In this case, as was shown earlier, \emph{Memory} is accessed exactly
once.
\item The RPC component may make a call to the reliable memory
component and pass the (exceptional) return value to the caller.
In this case, \emph{Memory} is never accessed.
\end{itemize}
Thus, every operation of the RPC component may result in zero or one
atomic reads of location $l$ in \emph{Memory}.  Consequently, 
the entire sequence of RPC component calls may result in zero or more
atomic reads of location $l$ in \emph{Memory}.  QED.

\begin{lem}
Every successful $\vbl{Read}(l)$ operation performs one or more
atomic reads to location $l$ at some time between the call and return.
\end{lem}

Note that this is different from the original requirement:
that a successful $\vbl{Read}(l)$ operation performs exactly one
atomic read of location $l$.  The described composition given above
cannot possibly satisfy this requirement.  To see this, observe that
the \emph{RPCFailure} exception can be raised by the RPC component
before any call to the Reliable Memory component (in which case no
read of $l$ occurs) or after it calls the Reliable Memory component
(in which case a single read of $l$ has occurred).  The implementation
component above cannot tell the difference between these two
conditions; since it is to retry the call some number of times 
before failing, we cannot ensure that a read to $l$ only occurs once
if retries are to be permitted.  We choose to allow retries and
proceed to prove the modified requirement.

\smallskip
\noindent \textbf{Proof.}  As in the previous lemma, the first
and last move in a successful $\vbl{Read}(l)$ operation are
the corresponding calls and return between the environment component
and the implementation component.  The moves between these two
elements can be divided into one or more disjoint subsequences of
moves; the last of these subsequences is a successful RPC operation,
while the remainder (if any) are unsuccessful RPC operations.

The previous lemma shows that a sequence of unsuccessful RPC
operations for a $\vbl{Read}(l)$ call results in zero or more
atomic reads to $l$.  A similar argument shows that a successful RPC
call results in a single atomic read to $l$; the result follows.  QED.

\begin{lem}
Every unsuccessful $\vbl{Write}(l,\,v)$ operation 
performs a sequence of zero or more atomic writes of
value $v$ to location $l$ at some time between the call and return.
\end{lem}

\begin{lem}
Every successful $\vbl{Write}(l,\,v)$ operation 
performs a sequence of one or more atomic writes of
value $v$ to location $l$ at some time between the call and return.
\end{lem}

The proof of these lemmas are similar to those for $\vbl{Read}$
operations and are thus omitted.  Combining these lemmas yields the
desired conclusion:

\begin{thm}
The ASM specification given 
correctly implements the requirements given for a memory component,
except that \vbl{Read} operations may perform more
than one atomic read between the call and return.
\end{thm}

%%%%%%%%%%%%%%%%%%%%%%%%%%%%%%%%%%%%%%%%%%%%%%%%%%%%%%%%%%%%%%%%%%%%%%

\section{A Lossy RPC Component}

The Broy-Lamport problem calls for the specification of
a Lossy RPC Component, whose requirements are as follows:

\begin{quotation}
The Lossy RPC component is the same as the RPC component except for
the following differences, where $\delta$ is a parameter.
\begin{itemize}
\item The \vbl{RPCFailure} exception is never raised.  Instead of
raising this exception, the \vbl{RemoteCall} procedure never 
returns.

\item If a call to \vbl{RemoteCall} raises a \vbl{BadCall} exception,
then that exception will be raised within $\delta$ seconds of the
call.

\item If a $\vbl{RemoteCall}(p,\,a)$ call results in a call of
procedure $p$, then that call of $p$ will occur within $\delta$
seconds of the call of \vbl{RemoteCall}.

\item If a $\vbl{RemoteCall}(p,\,a)$ call returns other than by
raising a \vbl{BadCall} exception, then that return will occur within
$\delta$ seconds of the return from the call to procedure $p$.
\end{itemize}

\begin{problem}
Write a formal specification of the Lossy RPC component.
\end{problem}
\end{quotation}

Clearly the requirements suggest the need for a modeling environment
which includes time.  We use the real-time ASM model
presented in \cite{railroad} and reviewed in Section 1.  

Implicit in the description above is the fact that every call and
return occurs at a specific moment in time.  Consequently, our
ASM descriptions of calls and returns will need to
record the time at which each call and return occurs.  
Our ASM will make use of several new unary functions:
\begin{itemize}
\item \emph{CallInTime}: the time that a call was received
by the given process 
\item \emph{CallOutTime}: the time that a call was placed
by the given process
\item \emph{ReturnTime}: the time that a return was received
by the given process
\end{itemize}
The new definitions for the CALL and RETURN abbreviations
are shown in Figure \ref{newcall}.

\BeginRule
\Abbreviation{CALL(procname, arglist, destination)}
   \>\Choose $p$ \Satisfying (Component(p)=destination \And CallSender(p)=undef)\\
   \>   \>CallSender(p) := Me\\
   \>   \>CallName(p) := procname\\
   \>   \>CallArgs(p) := arglist\\
   \>   \>CallInTime(p) := CT\\
   \>   \>ReturnTime(p) := undef\\
   \>   \>CallOutTime(Me) := CT\\
   \>   \>CallMade(Me) := true\\
   \>   \>CallReply(Me) := undef\\
   \>   \>CallReplyValue(Me) := undef\\
   \>\Endchoose\\
\\
\Abbreviation{RETURN(type, value)}
   \>CallReply(CallSender(Me)) := type\\
   \>CallReplyValue(CallSender(Me)) := value\\
   \>ReturnTime(CallSender(Me)) := CT\\
   \>CallOutTime(CallSender(Me)) := undef\\
   \>CallInTime(CallSender(Me)) := undef\\
   \>CallSender(Me) := undef\\
   \>CallName(Me) := undef\\
   \>CallArgs(Me) := undef\\
   \>CallMade(Me) := false\\
\EndRule{newcall}{The new CALL and RETURN abbreviations.}

The ASM program for the lossy RPC component is given
in Figure \ref{lossyrpcea}.  It uses a distinguished element $\delta$
as specified in the problem description.  

\BeginRule
\If CallName(Me) = remotecall \Then\\
   \>\If CallInTime(Me) $\neq$ undef \And CallOutTime(Me)=undef \Then\\
   \>   \>\If CT $\geq$ CallInTime(Me) $+ \delta$ \Then FAIL\\
   \>   \>\Elseif Length(Second(CallArgs(Me))) $\neq$ ArgNum(First(CallArgs(Me))) \Then\\
   \>   \>   \>RETURN(exception, BadCall)\\
   \>   \>\Else\\
   \>   \>   \>CALL(First(CallArgs(Me)),Second(CallArgs(Me)),Destination)\\
   \>   \>\Endif\\
   \>\Elseif ReturnTime(Me) $\neq$ undef \Then\\
   \>   \>\If CT $\geq$ ReturnTime(Me) $+ \delta$ \Then FAIL\\
   \>   \>\Else RETURN(CallReply(Me), CallReplyValue(Me))\\
   \>   \>\Endif\\
   \>\Endif\\
\Endif\\
\\
where FAIL abbreviates\\
   \>CallName(Me) := false\\
   \>CallArgs(Me) := false\\
   \>CallMade(Me) := false\\
   \>CallInTime(Me) := undef\\
   \>CallOutTime(Me) := undef\\
   \>ReturnTime(Me) := undef
\EndRule{lossyrpcea}{Lossy RPC component program.}

\begin{lem}
Every operation of the Lossy RPC component has one of the following
forms:
\begin{itemize}
\item A call which results in no call of the destination component
and no return to the caller
\item A call which results in a \emph{BadCall} exception being
raised within $\delta$ seconds of the call
\item A call which results in a call of the destination component
within $\delta$ seconds of the call, but results in no
return to the caller
\item A call which results in a call of the destination component
within $\delta$ seconds of the call, whose return is relayed to
the caller within $\delta$ seconds of the return
\end{itemize}
\end{lem}
This lemma can be easily verified by analysis of the program above.
The key point to notice is that any call or return action by
the Lossy RPC component is guaranteed to occur within $\delta$
seconds of the event which prompted that action; if too much
time passes, the rules ensure that the FAIL abbreviation will
be executed instead of the call or return action.

%%%%%%%%%%%%%%%%%%%%%%%%%%%%%%%%%%%%%%%%%%%%%%%%%%%%%%%%%%%%%%%%%%%%%%

\section{The RPC Implementation}

The Broy-Lamport calls for one final implementation, whose
requirements are as follows:

\begin{quotation}
The RPC component is implemented with a Lossy RPC component by passing
the \vbl{RemoteCall} call through to the Lossy RPC, passing the return
back to the caller, and raising an exception if the corresponding
return has not been issued after $2\delta+\epsilon$ seconds.

\begin{problem}
(a) Write a formal specification of this implementation.  

(b) Prove that, if every call to a procedure in \vbl{Procs} returns
within $\epsilon$ seconds, then the implementation satisfies the
specification of the RPC component in Problem~2.
\end{problem}
\end{quotation}

The implementation module is shown in Figure \ref{imp2ea}.  It uses a few new 
functions whose meaning should be clear by now.

\BeginRule
\If CallName(Me) $\neq$ undef \Then\\
   \>\If CallMade(Me) = false \Then\\
   \>   \>CALL(CallName(Me), CallArgs(Me), LossyRPC)\\
   \>\Elseif CallReply(Me) $\neq$ undef \And
	ReturnTime(Me) $\leq$ CallOutTime(Me) $+ 2\delta + \epsilon$ \Then\\
   \>   \>RETURN(CallReply(Me), CallReplyValue(Me))\\
   \>\Elseif (CT $\geq$ CallOutTime(Me) $+ 2\delta + \epsilon$) \Then\\
   \>   \>RETURN(exception, RPCFailure)\\
   \>\Endif\\
\Endif
\EndRule{imp2ea}{RPC implementation component module.}

We combine this implementation module with an instance of the
caller module, an instance of the lossy RPC module,
and an instance of the memory (or reliable memory) Component 
(so that the lossy RPC module has someone to whom it passes calls).
We assert that the memory component is bounded with bound $\epsilon$.

\begin{thm}
Every operation of the implementation component above has one of the
following forms:
\begin{itemize}
\item a call to the LossyRPC component which returns a \vbl{BadCall}
exception
\item a call to the LossyRPC component which makes no call
to the Memory component; the implementation component
then returns an \vbl{RPCFailure} exception 
\item a call to the LossyRPC component which makes a call
to the Memory component, waits for the return value from the
Memory component, and ignores the
return value; the implementation component then returns an
\vbl{RPCFailure} exception 
\item a call to the LossyRPC component which makes a call
to the Memory component from the Memory component, 
waits for the return value, and returns the value
\end{itemize}
\end{thm}

\noindent \textbf{Proof.} The previous lemma establishes the
behavior of the lossy RPC component, to which the implementation
component forwards its calls.  Notice that any operation which
does not result in an \vbl{BadCall} or an \vbl{RPCFailure} exception
requires an operation between the lossy RPC component and the
memory component, which by supposition is guaranteed to complete
within time $\epsilon$.  Notice that further that the lossy RPC
component must send its call to the memory component within $\delta$
seconds of the call in order to receive any return;
otherwise, the FAIL abbreviation will be executed, discarding the call.
Similarly, the lossy RPC component
must relay the return value from the Memory component to the
implementing component within $\delta$ seconds of the return
in order to return anything at all.  Thus, any successful return
will occur within $2\delta + \epsilon$ seconds of the original
call.  The implementing component can thus safely assume that any
call to the lossy RPC component which has not resulted in a 
return within that time interval will never have a return, and
the \vbl{RPCFailure} exception may be safely generated.  QED.

\end{document}